\documentclass[prd,showpacs,nofootinbib,twocolumn,amsmath,amsfonts,amssymb]{revtex4} 

\usepackage{amsmath}
\usepackage{amssymb}
\usepackage{epsfig}

\usepackage{graphicx}
\usepackage{stmaryrd}
\usepackage{bm}
\usepackage{color}

\newcommand{\eps}  {\epsilon}

\newcommand{\AddrCLUSTER}{%
 Excellence Cluster Universe, Technische Universit\"{a}t M\"{u}nchen,\\
 Boltzmannstra\ss e 2, D-85748, Garching, Germany}

\begin{document}
\title{Hint of non-standard dynamics in solar neutrino conversion}
\author{Antonio Palazzo}
\affiliation{\AddrCLUSTER}

\date{\today}

\begin{abstract}

Motivated by the recent low-threshold measurements  of the solar $^8$B neutrino spectrum performed by Borexino, Super-Kamiokande and the Sudbury Neutrino Observatory -- all now monitoring  the transition regime between low-energy (vacuum-like) and high-energy (matter-dominated) flavor conversions -- we consider the role of sub-dominant dynamical terms induced by new flavor-changing interactions.  We find that the presence of such perturbations with strength $\sim 10^{-1}G_{F}$ is now favored, offering a better description of the anomalous behavior suggested by the new results, whose spectrum shows no sign of the typical low-energy upturn predicted by the standard MSW mechanism. Our findings, if interpreted in a 2-flavor scheme, provide a hint of such new interactions at the $\sim 2 \sigma$ level,
which is rather robust with respect to 3-flavor effects possibly induced by non-zero $\theta_{13}$.

\end{abstract}


\maketitle

\section {Introduction}

One of the most important achievements in our understanding of neutrino properties is 
undoubtedly constituted by the (indirect) proof of the existence of 
matter effects in solar neutrino flavor conversion, as predicted by the Mikheev-Smirnov-Wolfenstein (MSW) 
mechanism~\cite{Wolfenstein:1977ue,smirnov}.
In fact, the explanation of the solar neutrino problem  
requires a peculiar energy dependence of the electron neutrino
survival probability ($P_{ee}$), which is elegantly provided by MSW transitions
in adiabatic regime~\cite{Adia}. At low energies ($E \lesssim1$ MeV) matter effects play a negligible role and 
a (averaged) vacuum-like behavior emerges, giving rise to $P_{ee} \sim 1/2$, in agreement with the
determinations of the Gallium  experiments SAGE~\cite{abdurashitov:2002nt}  
and GALLEX/GNO~\cite{Hampel:1998xg,Altmann:2005ix,kirsten2008retrospect}.
At high energies
($ E \sim 10$ MeV) matter effects dominate, leading to a much stronger signal
suppression ($P_{ee} ~\sim1/3$), as confirmed by the accurate boron neutrino ($^8$B $\nu$) 
measurements performed by SNO~\cite{ahmad-ahmed,Aharmim:2009gd,SNO,Aharmim:2008kc}
and Super-Kamiokande~\cite{fukuda,SK-I,Yang:2009hp}. This behavior is indirectly corroborated by the 
Chlorine~\cite{cleveland:1998nv} experiment,
whose total rate gets two (inseparable) contributions from both  low- and high-energy
neutrinos, and has recently received direct confirmation by the real time measurements performed
by Borexino~\cite{Arpesella:2008mt,Bellini:2008mr}, able to monitor separately both regimes.
Such a picture is now considered a standard framework thanks to the spectacular
results achieved by the reactor experiment KamLAND~\cite{kamland,Gando:2010aa}, 
which has provided  the unique opportunity to measure the solar mass-mixing parameters
in vacuum, independently of the dynamical effects induced by neutrino interactions with matter. 

It is well known~\cite{Wolfenstein:1977ue,Valle:1987gv, guzzo:1991hi, roulet:1991sm} 
that non-standard neutrino interactions (NSI), described by low-energy four-fermion operators
$\mathcal{O}_{\alpha\beta}\sim \overline\nu_\alpha\nu_\beta \overline f f$
of sub-weak strength $\eps_{\alpha\beta}G_F$, 
especially of the flavor-changing type ($\alpha \ne \beta$), 
can profoundly modify the flavor conversion process. 
Interestingly, such new interactions may 
produce appreciable deviations {\em only}
in the intermediate energy region~\cite{Friedland:2004pp} describing the transition between 
vacuum-like and matter-dominated conversions, without affecting the well established behaviors observed
at low and high energies. Therefore, the accurate observation of this energy region
is of crucial importance for pinning down potential new physics beyond the Standard Model.

Intriguingly, the first low-threshold $^8$B $\nu$ measurements performed by Borexino~\cite{Bellini:2008mr}
and  by the SNO low energy threshold analysis~\cite{Aharmim:2009gd} (LETA), together with 
those provided by the older (SK-I~\cite{SK-I}) and newer (SK-III~\cite{Yang:2009hp}) 
Super-Kamiokande data, 
point towards an anomalous behavior, showing no evidence of
the low-energy upturn of the spectrum predicted by the standard MSW mechanism.
This new circumstance suggests that new interactions may be effectively at work, affecting the
conversion  of solar neutrinos in an observable way. In this communication we quantify such an expectation by showing 
that, with the inclusion of  the new spectral information, the solar
sector data (solar+KamLAND) display a non-negligible preference for NSI, 
disfavoring the standard MSW picture at the $\sim 2 \sigma$ level.

\section {Notation}
The evolution of a two neutrino system 
is governed, in the flavor basis, by a Schroedinger-like equation
\begin{equation}
\label{eq:2nuevol}
 i\, \frac{d}{dx}\left(\begin{array}{c}\nu_e\\ \nu_a \end{array}\right) = H
 \left(\begin{array}{c}\nu_e\\ \nu_a \end{array}\right)\ ,
\end{equation}
where $\nu_a$ is a linear combination of $\nu_\mu$ and $\nu_\tau$,
and  $H$ is the total Hamiltonian
\begin{equation}
H = H_\mathrm{kin} + H_\mathrm{dyn}^\mathrm{std} + H_\mathrm{dyn}^\mathrm{NSI}  
\label{eq:H_tot}\,,
\end{equation}
split in the sum of the kinetic term, the standard MSW term~\cite{Wolfenstein:1977ue,smirnov}, 
and of a new NSI-induced term. The kinetic term reads $H_\mathrm{kin} = UKU^T$,
where U is the real orthogonal $2\times2$ mixing matrix depending on the  mixing angle
$\theta_{12}$, and $K$ is the diagonal matrix of wavenumbers $k_i = m_i^2/2E$
($m_i$ and $E$ being the neutrino squared masses and energy respectively).
In the presence of ordinary matter, the standard electroweak theory predicts
$H^\mathrm{std}_\mathrm{dyn}=\mathrm{diag}(V,\,0)$, where $V(x) =\sqrt 2 G_F N_e(x)$
is the effective potential induced by the charged current $\nu_e$ interaction with the electrons
having number density $N_e(x)$. For interactions with a background 
fermion $f $ with number density $N_f(x)$, the new term  can be expressed as~\cite{Guzzo:2001mi}
\begin{equation}
    H^{\mathrm{NSI}}_{\mathrm{dyn}} =
       \sqrt 2 G_F N_f(x)
    \begin{pmatrix}
        0         & \eps \\
        \eps  & \eps'
    \end{pmatrix}
\,,    
\end{equation}
where $\eps$ and $\eps'$  are two effective parameters which, 
restricting to the case of flavor-changing interactions with $d$-quarks, 
are related to the fundamental vectorial couplings $\eps_{\alpha\beta}^{dV}$
as~\cite{Guzzo:2001mi}
\begin{align}
    \label{eq:4a} \eps
    &=  \eps_{e\mu}^{dV}  \cos \theta_{23} - \eps_{e\tau}^{dV}  \sin\theta_{23} \ ,
    \\
 \begin{split}
        \label{eq:5a} \eps'
        &= - \eps_{\mu\tau}^{dV} \sin 2 \theta_{23}  \, 
       \end{split} \,,
\end{align}
where $\theta_{23}$  is the atmospheric mixing angle. 
Taking into account the strong upper bounds on $\eps_{\mu\tau}^{dV}$
deriving from the atmospheric data analysis~\cite{Fornengo:2001pm,GonzalezGarcia:2007ib},
we can safely neglect the diagonal effective coupling $\eps'$.
Therefore, the conversion of solar neutrinos is described
by the mass-squared  splitting  $\Delta m^2 = m_2^2 - m_1^2 $ ,
the mixing angle $\theta_{12}$, and the effective parameter $\eps$.

\begin{figure}[b!]
\vspace*{-4.0cm}
\hspace*{-0.4cm}
\includegraphics[width=13.0 cm]{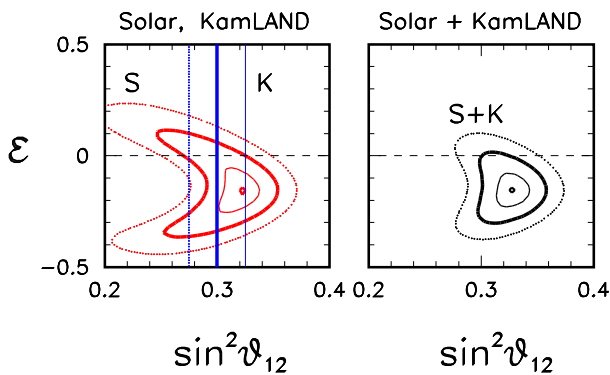}
\vspace*{-4.0cm}
\caption{Region allowed, after marginalization of $\Delta m^2$ as constrained by KamLAND, separately (left panel)
by solar (S) and KamLAND (K) data and by their combination (right panel).
The contours refer to $\Delta \chi^2 =1$ (thin solid line), $\Delta \chi^2 = 4$ (thick solid line),
and $\Delta \chi^2 = 9$ (dotted line).
\label{fig2}}
\end{figure}  
\begin{figure}[t!]
\vspace*{-1cm}
\hspace*{-0.2cm}
\includegraphics[width=7cm]{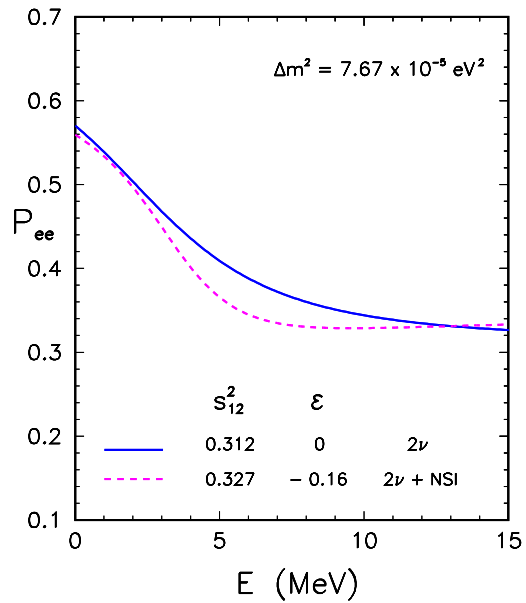}
\vspace*{-0.2cm}
\caption{Solar $\nu_e$ survival probability (averaged over the $^8$B $\nu$
production region) for the best fit points obtained with (dashed line)
and without (solid line) NSI effects.}
\end{figure}  

\section {Numerical results}

In our analysis we have included the  
data from Homestake~\cite{cleveland:1998nv},
SAGE~\cite{abdurashitov:2002nt} and
GALLEX/GNO~\cite{Hampel:1998xg,Altmann:2005ix,kirsten2008retrospect},
SK-I~\cite{SK-I},
the third SNO phase~\cite{Aharmim:2008kc},
and the Borexino $^7$Be data~\cite{Arpesella:2008mt}. In addition, we have incorporated
the new spectral information provided by SNO-LETA~\cite{Aharmim:2009gd}, 
SK-III~\cite{Yang:2009hp},  and  the $^8$B  Borexino data~\cite{Bellini:2008mr}.
We have also included the latest KamLAND results~\cite{Gando:2010aa}.
For the sake of precision, we have incorporated both standard and non-standard matter effects also
in the KamLAND analysis. However, due to the low density of the Earth's crust,
both have only a negligible impact for the range of parameters we are
considering.  Therefore,  the constraints obtained
from KamLAND do not depend on NSI. We also included NSI effects in the
propagation of solar neutrinos in the Earth which, as noted
in~\cite{Friedland:2004pp}, can slightly modify the regeneration effect.

\begin{figure*}[t!]
\vspace*{-5.5cm}
\hspace*{1.4cm}
\includegraphics[width=20.0 cm]{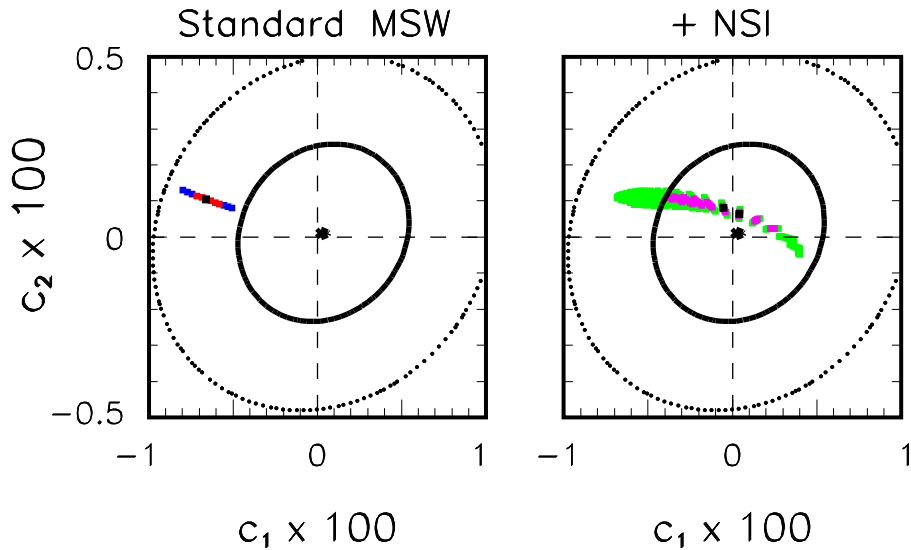}
\vspace*{-7.3cm}
\caption{The (ellipse-like) region determined by all data sensitive
to $^8$B neutrinos is reported in both panels. In the left (right) panel the
region spanned by the theoretical model in the absence (presence) of NSI
effects is superimposed. Both experimental and theoretical
regions are plotted for  $\Delta \chi^2 = 1,4$.}
\end{figure*}  

In Fig.~1, we display the results of the analysis, by showing
the allowed regions 
in the plane charted by [$\sin^2\theta_{12}, \eps$],
after marginalization of $\Delta m^2$,  which in practice
is fixed by KamLAND. In the first panel we show separately the region 
determined by solar and KamLAND data. As already noticed in~\cite{Palazzo:2009rb}, 
for non-zero (negative) values of $\eps$ the solar data tend to prefer larger values 
of $\theta_{12}$, with an improved agreement with KamLAND.
We also notice that the solar data taken {\em alone} tend to prefer non-zero NSI. 
As we will discuss below, this preference can be traced to 
the anomalous behavior of the solar energy spectrum  suggested by the present data.
In the right panel we report the region allowed by the combination
of solar and KamAND data. This plot shows that the standard
MSW case ($\eps = 0$) is disfavored almost at the $2 \sigma$ level; 
more exactly we find $\Delta \chi^2 \simeq 3.6$, corresponding to $1.9\sigma$. 
The best fit is obtained%
\footnote{Interestingly, a similar preference for flavor-changing NSI
in the ($\nu_\mu,\nu_\tau$) sector has been evidenced in
connection with the mismatch of the atmospheric mass-squared splittings
observed, respectively, in the neutrino and antineutrino measurements performed by MINOS~\cite{Mann:2010jz,Kopp:2010qt}.} 
for $\eps \simeq -0.16$, which, assuming maximal atmospheric mixing~\cite{Fogli:2008ig}, corresponds to 
a difference  $\eps_{e\tau}^{dV} - \eps_{e\mu}^{dV} =  0.23$, well
compatible with the existing experimental bounds~\cite{limits}.

For definiteness, we have focused on the case of interactions with d-type quarks.
However, the essence of our results is unaltered if interactions with u-type quarks or
electrons are considered. In all cases the absence of non-standard effects
is disfavored at the same statistical level.  What changes is the best-fit value of $\eps$, 
as can be expected from the proportionality of the new dynamical term to the fermion 
number density  [see eq. (3)].

\section {Discussion}
In Fig.~2 we show the solar $\nu_e$ survival
probability (averaged over the $^8$B $\nu$ production region) as a function of the neutrino energy,
for the best fit points obtained with or without NSI, respectively. 
In both cases the mass squared splitting is $\Delta m^2 = 7.67\times10^{-5} \mathrm{eV}^2$,
while the best fit values of  $\theta_{12}$ are slightly
different, as can be understood from the second panel of Fig.~1.
It is evident how in the presence of NSI, the survival probability exhibits
a distinctive flat behavior for  $E\gtrsim7$~MeV, in contrast with the standard MSW case
which, in the same energy range, presents a negative slope.

To clarify the role of the spectral information in favoring the flatter NSI solution,
following~\cite{Aharmim:2009gd}, we parametrize the survival probability averaged over
the full day (daytime and nighttime%
\footnote{Hereafter, we neglect the small differences in the (nighttime)
survival probability probed at the three detectors Borexino, SNO and SK,
induced by their different latitudes.  We have checked that this approximation does
not alter our conclusions.}) as a second order polynomial  
\begin{equation}
 P_{ee}(E) = c_0 + c_1 (E-E_0) + c_2 (E-E_0)^2
 \end{equation}
where $E_0 = 10$~MeV approximately represents the
energy where the $^8$B $\nu$ experiments
are most sensitive. We have extracted the three coefficients $c_i$
from the combination of all the high energy
experiments, taking into account their sensitivities.
For SNO LETA this extraction is not feasible, as
detailed information on the bin-to-bin correlations of the spectrum
is not available, and we have directly used the coefficients' information
provided by the collaboration itself~\cite{Aharmim:2009gd}.
 
In both panels of Fig.~3 we report the (ellipse-like) region obtained by projecting
the 3-dimensional allowed region onto the 2-dimensional parameter
space of  the two coefficients ($c_1, c_2$), which encode the
deviations from flatness. The contours correspond to the $1\sigma$ (continuous
curve)  and $2\sigma$ (dotted curve) level.  The best fit coincides with the origin
of the axes signaling a clear preference for a flat spectrum. In the left panel 
we superimpose the region spanned in the space of these two coefficients
by the theoretical model in the absence of NSI. 
From this plot we learn that the standard MSW mechanism corresponds to a very well definite region  
in this plane:  the variation of $\theta_{12}$ in the range allowed by the data
produces only a modest excursion from the best fit point,  which is obtained
for a negative value of $c_1$, corresponding to the expected low-energy 
upturn of the survival probability.  The value  of the experimental $\Delta \chi^2 \sim 2.1$ 
assumed in this point provides a quantitative measure
of the slight disagreement among theory and data. In the right panel we superimpose the region 
determined in the presence of NSI. In this case, the theoretical  best fit point
lies almost exactly at the center of the ellipse-like region,  thus offering a perfect
description of the flat spectrum indicated by the current data. Therefore, the NSI scenario, 
respect to the standard MSW case, ``gains''  a negative $\Delta \chi^2 \sim - 2.0$, 
which can be identified as the partial contribution arising from the 
better description of the energy spectrum to the total
value $\Delta \chi^2 \sim 3.6$ emerging from the global fit.

\begin{figure}[t!]
\vspace*{-2.6cm}
\hspace*{-0.6cm}
\includegraphics[width=13.0 cm]{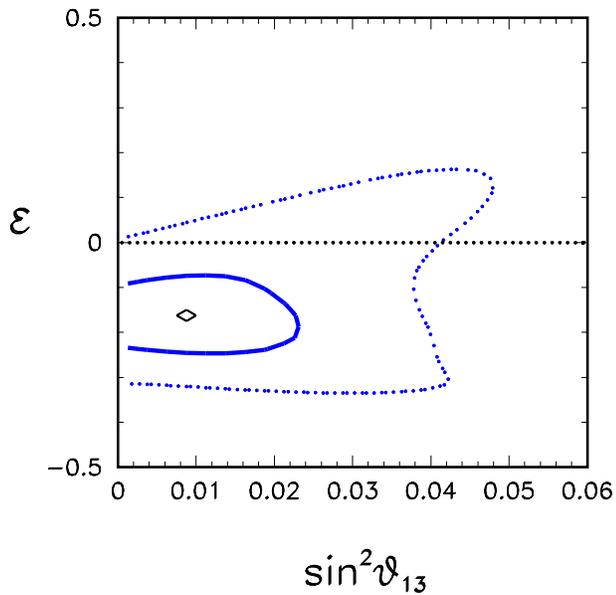}
\vspace*{-2.5cm}
\caption{Region allowed, after marginalization of $\Delta m^2$ and $\theta_{12}$, 
by the combination of solar and KamLAND data. The contours refer to $\Delta \chi^2 =1$ 
(solid line) and $\Delta \chi^2 = 4$ (dotted line).
\label{fig4}}
\end{figure}  

The remaining $\Delta \chi^2 \sim - 1.6$ can be traced~\cite{Palazzo:2009rb} to the better agreement
obtained in the presence of NSI, among the slightly different values of the
mixing angle $\theta_{12}$  determined, respectively, by solar and KamLAND data
(see Fig.~1). As discussed in~\cite{Fogli:2008jx,Balantekin:2008zm,Palazzo:2009rb} this 
slight tension can be equally alleviated by the kinematical 
effects induced by non-zero values of the third mixing angle $\theta_{13}$.
It must be noted, however, that the observed flat behavior of the spectrum cannot  be reproduced by
a non-zero value of $\theta_{13}$, since this parameter induces  only an energy-independent
suppression of the 2-flavor survival probability~\cite{Palazzo:2009rb}. Therefore, in the general three-flavor case
we expect only a modest reduction of the statistical preference for non-zero NSI.
The robustness of the hint is confirmed by the 3-flavor analysis, whose results 
are shown in Fig.~4, reporting the allowed region in the plane [$\sin^2 \theta_{13}, \eps$],
having marginalized away all the other parameters. Allowing for $\theta_{13}>0$,
the analysis still indicates a preference for the new effects at the $1.5\sigma$ level. The future 
reactor searches~\cite{reactor} will provide a measurement of $\theta_{13}$ unaffected
by (standard and non-standard) dynamical effects. Therefore, their negative (positive) result,
will slightly favor (disfavor) the NSI hypothesis discussed here.

\section {Conclusions}

We have shown that the latest solar neutrino data (in combination
with KamLAND) favor the presence of non-standard dynamical terms in the MSW Hamiltonian
at a non negligible statistical level, thus hinting at new neutrino interactions. We have shown
how such an indication, already present with lower statistical significance in the older 
data~\cite{Palazzo:2009rb,Escrihuela:2009up}, is now
enhanced by the anomalous spectral behavior observed in three experiments.  We stress that the
indication we have discussed is indirect, and may be confused with other possible sources of anomalous 
spectral distortions, as those induced by conversions into new sterile neutrino 
states~\cite{sterile}.
Therefore, the identification of the correct sub-leading effect (if any) will need 
further corroboration not only from new indispensable low-energy solar neutrino
measurements,  but also from all the remaining neutrino  phenomenology.

\section*{Acknowledgments}
We would like to thank A. Ianni for useful information on the Borexino detector. 
Our work  is supported by the DFG Cluster of Excellence `Origin and Structure of the Universe'.

\bibliographystyle{h-physrev4}

\end{document}